\documentstyle[11pt,paspconf,epsf]{article}
\input{psfig.tex}

\begin{document}

\title{The power of jets: blazars vs galactic superluminals}

\author{Gabriele Ghisellini}
\affil{Osservatorio Astronomico di Brera, V. Bianchi 46, I-23807 Merate Italy}

\begin{abstract}
Estimates on different scales of the power of relativistic 
bulk motion in extragalactic and galactic jets are presented.
The power in the jets and the power produced by the accretion 
disk are found to be roughly equal.
This may suggest an important role of the magnetic field in extracting
both the gravitational energy of the accreting matter and the 
rotational energy of the black hole.
A method to derive the minimum power of jets is discussed,
and used to find lower limits to the jet power of blazars
and the galactic superluminal source GRS 1915+105.
The matter content of jets is discussed, finding
that electron positron pairs cannot be dynamically important.
The jet power can initially be in the form of Poynting flux,
gradually accelerating matter until equipartition between
bulk motion of the particles and Poynting flux is reached.
\end{abstract}

\keywords{Jets, AGNs, blazars, galactic superluminals, radiation processes:
synchrotron, inverse Compton, electron--positron pairs}

\section{Introduction}
Special relativity is important to study fundamental 
particles in terrestrial accelerators, but it is also the key 
to understand large scale phenomena involving thousand of solar 
masses traveling at nearly the speed of light for kpc distances.
Extragalactic jets, as well as their galactic superluminal counterparts,
may well carry more power than what is emitted by their accretion disk.
Despite the prediction that jet carry plasma in relativistic motion
dates back to 1966 (Rees, 1966), and intense studies over the last
20 years (Begelman, Blandford \& Rees, 1984),
quantitative estimates of the amount 
of power that jets can transport have been done only relatively recently,
following new observational results.
The  main difficulty for a quantitative analysis is that the amount
of observed power emitted by jets is strongly modified 
by relativistic aberration, time contraction and blueshift, 
all dependent on the unknown plasma bulk velocity and viewing angle.
The other fundamental unknown is the matter content of jets:
we still do not know for sure if they are made by electron--positron
pairs or normal electron--proton plasma.

However, there have been many attempts to calculate the power of jets,
both in the form of bulk kinetic power and in intrinsic emitted luminosity.
In this paper I will summarize some of the recent studies aiming
to know the power of jets, and to put some constraints on the jet content.
Although I will focus mainly on AGN jets,
many of the arguments presented below can be applied both to 
extragalactic as well galactic superluminal jets, which will
be briefly reviewed in the end.

\section{The lobes of strong extragalactic radio sources}

The fact that the lobes of strong radio sources contain a huge amount 
of energy was known for many years, but 
Rawlings and Saunders (1991) put the accent on the $power$
that the lobes require to exist at all.
Their argument is conceptually simple, since they derive the power
dividing the total energy by the lobe life timescale.
This is the power that jets must supply.
This important result depends upon the assumption of
equipartition between particles and magnetic field in the lobes,
while the life timescale was derived with 
spectral aging of the observed synchrotron emission and/or by
dynamical arguments. 
When both estimates were possible, they were found to agree.
In addition, they found that the derived power correlates with the amount
of luminosity observed in the narrow emission lines, which are another
isotropic indicator of power, but probably coming from a different origin.
The found correlation (once both low and high luminosity objects are included)
spans more than 4 decades of power on both axis, and it is more informative
than the previously found correlation between the lobe radio power 
and the narrow line luminosity.
This is because only a small fraction of the power received by the lobes
may be dissipated into radiation.

\section{The jet at VLBI scale}

Interferometry can sample sizes of a few pc for objects
located at a redshift $z=1$.
This has allowed us to discover superluminal motions of knots in the jets,
the main proof of relativistic bulk motion.
Radio size and flux constrain the beaming factor
of the radio emission, in order not to exceed, by the inverse Compton
process, the observed amount of X--rays.
If we then restrict ourselves to blazars, believed to be seen at small
viewing angles, we can estimate the bulk Lorentz factor $\Gamma$,
or directly estimate it $and$ the viewing angle by adding the information
of the apparent velocity $\beta_{app}$. 
The main uncertainties in these estimates are
the size of the sources, often at the limit of the instruments, and
the Hubble constant ($\beta_{app}\propto 1/H_0$).
Therefore average results, coming from a large sample of objects
(Ghisellini et al. 1993)
are more reliable than quantities derived for a specific object.
If the core radio flux is emitted by the incoherent synchrotron 
process, we can estimate the magnetic field, and the number 
density $n$ of the
emitting particles, which however depends crucially by the low energy
cut--off ($\gamma_{min}$) of their energy distribution, where most
of the particles are.
Since low energy particles would emit self--absorbed synchrotron emission, 
$\gamma_{min}$ cannot be estimated by the radio data.
Bearing these uncertainties in mind, 
Celotti \& Fabian (1992) calculated the bulk motion kinetic power
of the jet at VLBI scales:

\begin{equation}
L_{k, VLBI}\, = 
\, \pi R_{VLBI}^2 \Gamma^2\beta n^\prime m_ec^3 (<\gamma> +a m_p/m_e)
\end{equation}
where $n^\prime =n/\Gamma$ is the intrinsic density of particles,
which have a mean energy $<\gamma>m_ec^2$.
$a$ measures the amount of protons per electron ($a=1$ corresponds to
equal number of protons and electrons). 
Protons are assumed to be cold.
The calculated $L_k$ was found to match the lobe power if:
i) the jet is composed by pairs, with $\gamma_{min}\sim 1$; or
ii) the jet is made by protons and electrons, but in this
case $\gamma_{min}\sim 30$--100.
Celotti \& Fabian (1992) preferred the latter solution, since otherwise
the required density of pairs, extrapolated from the VLBI
scale back to the initial part of the jet cannot survive against
annihilation (Ghisellini et al. 1992).
Note that a low energy cut--off of the same order (in the case of 
proton--electron plasma) is required for polarized sources, 
to limit the amount of Faraday depolarization.
Another important result is that there must be a little amount
of thermal particles, since they would  increase too much both
the bulk kinetic power and the Faraday depolarization
(Wardle 1977).

\begin{figure}
\psfig{file=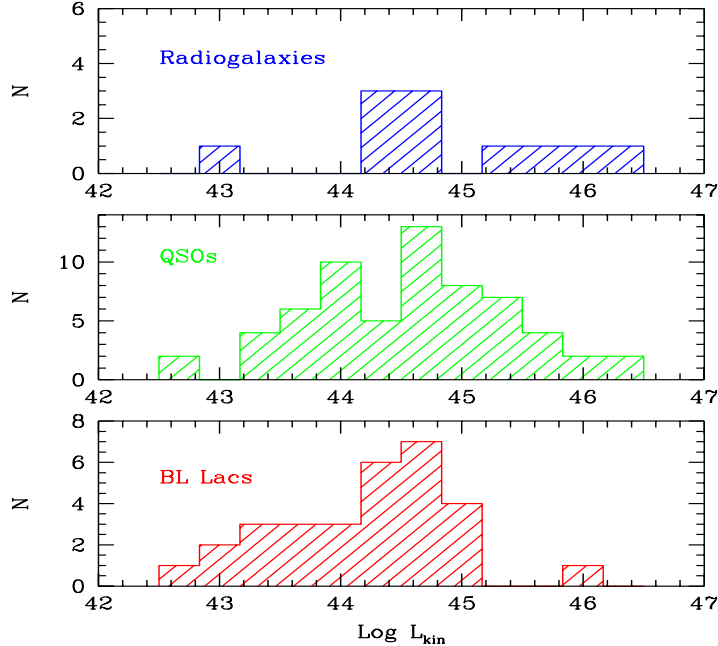,width=11truecm,height=9truecm}
\caption[h]{Bulk kinetic power for radio-galaxies, quasars and BL Lacs
as calculated in Celotti \& Fabian 1993. $\gamma_{min}=10$ and
an electron--proton jet is assumed.}
\end{figure}

\section{Jet power vs accretion disk luminosity}

The kinetic power of jets can be best estimated for blazars, since
for these sources is much easier to limit the bulk Lorentz factor.
On the other hand in these sources the beamed continuum hides
the isotropic continuum component at all frequencies but the radio ones.
Comparing the bulk kinetic power with the nuclear luminosity 
produced by accretion is therefore difficult.
One way out is to use emission lines, assuming 
that they reprocess a relatively fixed fraction of
the ionizing accretion disk luminosity.
In this way Rawlings and Saunders (1991) found  
$<L_{NLR}/L_{lobe}>\sim 0.01$,
while Celotti, Padovani \& Ghisellini (1997) found
$<L_{BLR}/L_{k, VLBI}>\sim 0.1$. 
NLR and BLR stand for narrow and broad emission line regions, respectively.
Note that BL Lac objects, lacking emission lines, do not obey
these correlations (but FR I radio sources do, Rawlings \& Saunders 1991).

Since $L_{NLR}\sim 0.1 L_{BLR}\sim 0.1 L_{ioniz} \sim L_{disk}$ and
since $L_{k, VLBI}\sim L_{lobe}\sim L_{jet}$
we arrive at the very important conclusion that
\begin{equation}
L_{jet} \sim L_{disk}
\end{equation}

\noindent
Alternatively, one may assume that a significant fraction of the
line regions is illuminated by the beamed jet emission, providing
a link between bulk motion power and line luminosity.
But in this case steep radio sources should have optical spectra with
huge line equivalent widths (these objects are seen at large viewing
angles, and the non-thermal continuum is depressed), contrary
to what we observe.

We do not know yet the mechanism responsible for the formation
and the acceleration of jets. 
Early attempts to use the strong radiation pressure of the accretion
disk failed, due to the strong Compton drag that the moving plasma
suffers even at modest speeds.
An attractive idea, but not yet explored in detail, is the Blandford
\& Znajek (1977) process, in which the magnetic field can extract the
rotational energy of a Kerr black hole:
\begin{equation}
L_{rot}\, \sim \, 10^{45} \left({ a\over m}\right)^2 M_8^2 B_4^2 \,\,\,\,
{\rm erg~s ^{-1}}\, \,  \sim \,
\left({ a\over m}\right)^2  (3R_s)^2 U_B c
\end{equation}
where the magnetic field $B=10^4 B_4$ Gauss, $U_B=B^2/(8\pi)$,
the black hole mass
$M=10^8 M_8$ solar masses, $R_s$ is the Schwarzchild radius and $(a/m)$
is the specific black hole angular momentum ($\sim 1$ for maximally rotating
black holes).
The accretion disk luminosity can be written as

\begin{equation}
L_{disk} \, =\, \pi (3R_s)^2 U_r c
\end{equation}
where $U_r$ is the energy density of the radiation produced by the disk.
For $U_B=U_r$ we have $L_{disk}\sim L_{rot}$ (Celotti et al. 1997).
These very simple estimates may be a coincidence, or may testify
the crucial role of the magnetic field on AGNs, responsible for
both transforming gravitational energy into radiation in the disk
and by extracting rotational energy from the black hole.

\subsection{Outflowing mass}

The jet and the accretion powers can be expressed as

\begin{equation}
L_k \, =\, \Gamma \dot M_{out} c^2; \qquad 
L_{disk} \, =\, \eta \dot M_{in} c^2
\end{equation}
where $\dot M_{in}$ is the mass accretion rate, and $\eta$
is the efficiency of mass to energy conversion for accretion.
Therefore we have 
\begin{equation}
\dot M_{out} \, =\, {\eta \over \Gamma}\, { L_k \over L_{disk}} \dot M_{in}
\, \sim \,\, 0.01 \dot  M_{in}
\end{equation}
The outflowing mass is then a small fraction of the accreted one,
but it still corresponds to $\sim$0.1 solar masses per year for the most
powerful blazars.

\section{The $\gamma$--ray zone}
The discovery that blazars emit a significant (and dominant, during flares)
fraction of their power in the $\gamma$--ray band poses important
constraints on the part of jet responsible for this emission.

\begin{figure}
\vskip -1 true cm
\psfig{file=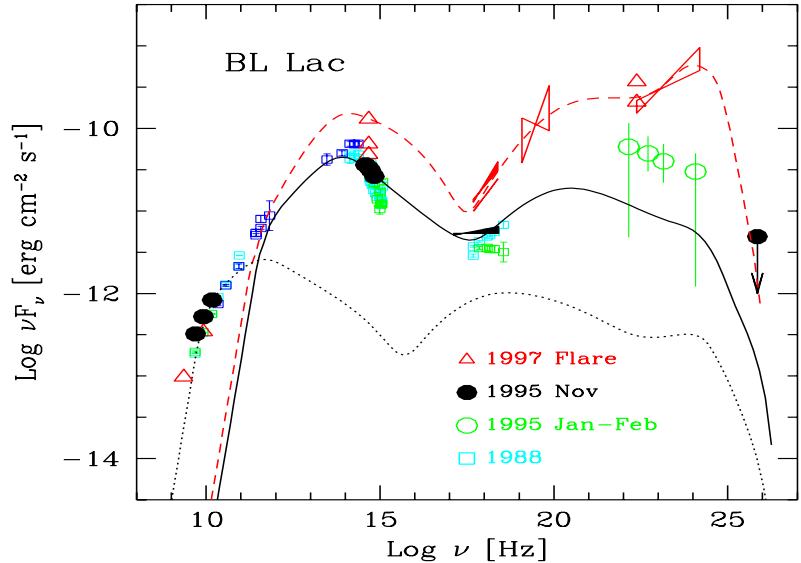,width=12truecm,height=9truecm}
\vskip -1 true cm
\caption[h]{Spectrum of BL Lac during different epochs. Lines
are fits with a synchrotron and inverse Compton model, including self Compton
and scattering with photons produced externally to the jet.
Note the large variations on the spectrum, and the change in the location
of the high energy peak. From Sambruna et al. (1998).}
\end{figure}

\begin{itemize}
\item The overall spectral energy distribution (SED) of blazars consists
of two broad peaks, one at frequencies between the IR and the soft X--rays
and the other one between the MeV and the GeV band.
Due to the limited EGRET sensitivity, we probably detect sources
in the $\gamma$--ray band when they flare, but a couple of
blazars (3C 279 and PKS 0528+134) were always detected when observed.
They suggest that the $\gamma$--ray power is of the same order 
than the power in the low energy peak during quiescent states,
and can be 10 or 100 times larger during flares.

\item The $\gamma$--ray flux varies rapidly. EGRET, onboard CGRO, detected
variations of a factor 2 in timescales of days.
Ground based Cerenkov telescopes have observed factor 2 flux
variations in a timescale as short as 20 minutes in Mkn 421
(Gaidos et al. 1996).
When possible, variations in the $\gamma$--ray band were observed 
to be accompanied by variations in other frequency bands, 
(Maraschi et al. 1994; Hartman et al. 1996;  Bloom et al. 1997; 
Wehrle et al. 1998; Buckley et al. 1996; Macomb et al. 1995).

These potentially very important observations for constraining
the different emission models suggest that the bulk of the emission,
both in the low energy peak, produced by synchrotron, and in the
high energy peak, presumably produced by inverse Compton, are cospatial
and produced by the same electrons.

\item If the X and the $\gamma$--ray emission are cospatial, then
the very fact to observe $\gamma$--rays implies bulk motion, to
evade the limit posed by the $\gamma$--$\gamma \to e^\pm$ process.
In fact this process depends of the compactness of the sources,
which is the intrinsic luminosity over size ratio.
Radiation must be beamed, in order to derive an intrinsic compactness
lower enough to let the source be transparent to $\gamma$--rays.
Lower limits to the beaming factors are consistent with those
derived by superluminal motion and the requirement not to overproduce
X--rays (Dondi \& Ghisellini 1995). 
These limits are particularly severe for TeV sources, since the optical
depth for $\gamma$--$\gamma$ collisions increases with energy
(for instance $\delta>$8--10 for Mkn 421), contrary to the old idea
that X--ray selected BL Lacs have smaller beaming factors,
being seen at larger viewing angles (Maraschi et al. 1986).

\item The above point constrains the location of the emitting region.
It must be at some distance from any important source of X--rays, which
would otherwise absorb the observed $\gamma$--rays.
The idea that $\gamma$--rays are produced all along the jet, and that 
only those produced at larger distances survive and arrive to us 
(Blandford \& Levinson 1994) must face the problem of reprocessing:
if a significant fraction of the power emitted in the inner part
of the jet in $\gamma$--rays get absorbed, the created pairs,
born relativistic in a dense photon environment, emit X--rays by
Compton scattering UV accretion disk photons, and inevitably the power
originally in the $\gamma$--ray band gets reprocessed into the X--ray band
(Ghisellini \& Madau 1996). 
The observed X--ray and $\gamma$--ray luminosities should therefore be roughly
equal, contrary to what observed.
There must be a mechanism able to transport energy from the central
powerhouse to a few hundreds Schwarzchild radii without dissipating it.

\item  The $\gamma$--ray emitting zone is where most of the power
is radiated away. 
Models interpreting the overall emission of blazars can take advantage 
of that, and assume an homogeneous region, rather than a many--parameter
inhomogeneous jet.
The location of the emitting region is constrained on one side by the 
requirement of $\gamma$--$\gamma$ transparency, and on the other side by 
the rapid variability: therefore 
it must be at some 10$^{17}$ cm from the black hole, and have a dimension
of $\sim 1/\Gamma$ times smaller.

\item It is interesting to compare, for this region, the power that
it must be transported in the form of bulk kinetic motion of the
plasma and the power in the form of radiation.
To this end, one needs to estimate the intrinsic parameters of the 
source, such as the value of the magnetic field, the particle density,
and the matter composition.
This has been done by Ghisellini et al. (1998) for all EGRET detected sources.
Note that these fits provide useful constraints for classes of sources,
rather than for specific objects, since the radio to $\gamma$--ray 
data are very rarely taken simultaneously, and since for few sources we have
a very good spectral coverage.

\end{itemize}

\section{The power of jets at the 0.1 pc scale}
Eq. (1), once $R_{VLBI}$ is replaced by the dimension estimated
by model fitting and by the variability timescale, measures
the bulk kinetic power of the jets at the 0.1 pc scale.
There is however another form of power: the Poynting flux:
\begin{equation}
L_B\, =\, \pi R^2 \Gamma^2 U_B c
\end{equation}
where $U_B$ is the intrinsic (comoving) magnetic energy density.
The observed synchrotron luminosity $L_{s,obs}$ is derived, assuming
isotropy, by multiplying the observed flux by $4\pi d_L^2$, where
$d_L$ is the luminosity distance.
Then $L_{s,obs}=\delta^4 L_s^\prime$.
Instead, the total synchrotron power received on the entire solid angle
is $L_{s,tot} =\delta^2 L_s^\prime$.
For a blob of dimension $R$ we have

\begin{equation}
L_{s,obs} \, =\, \delta^4 Vol\int n^\prime(\gamma)\dot\gamma_s m_ec^2d\gamma\, =
\, {16\pi R^3 \over 9} \delta^4 
\sigma_T c n^\prime U_B <\gamma^2>
\end{equation}
where $<\gamma^2>$ is averaged over the emitting particle distribution.
By substituting the particle density derived by this equation into Eq. (1),
and assuming $\delta=\Gamma$ we have
\begin{equation}
L_k\, =\, {9\pi m_ec^2 \over 2\sigma_T}\, 
{L_{s,obs} \over R \Gamma^2 B^2}
\, {<\gamma> + m_p/m_e \over <\gamma^2>}
\end{equation}
We see that $L_k \propto (B\Gamma)^{-2}$, while $L_B\propto (B\Gamma)^2$:
therefore $L_{jet}\equiv L_k+L_B$ is minimized for some value of 
$B\Gamma$, which corresponds
to equipartition between particle and magnetic energy densities.
In this case the $observed$ (at a viewing angle $\sim 1/\Gamma$)
synchrotron emission is maximized.
EGRET sources obey this condition if $\gamma_{min}$ (entering in the 
definition of $<\gamma>$ and $<\gamma^2>$) is again of the order of
$\sim 30$ and if electron--positron pairs are not dynamically important.
In addition, for this value of $\gamma_{min}$ we find particle
conservation between the $\gamma$--ray and the VLBI emitting zones
of the jet.

\subsection{Meaning of equipartition of power}

We are used to the equipartition argument regarding particle and
magnetic $energy$.
In the previous section we have instead apply the similar, but
conceptually different concept of equipartition of particle bulk
motion $power$ and Poynting flux.
The synchrotron power then relates the two.
We are then comparing the {\it velocity at which particle bulk kinetic
and magnetic energies are converted into radiation}.
With the assumption that the viewing angle is $1/\Gamma$ we
have that for a given observed synchrotron power, equipartition
between $L_k$ and $L_B$ corresponds to the minimum $L_{jet}=L_k+L_B$.
Alternatively, for a given $L_{jet}$, equipartition between
$L_k$ and $L_B$ ensures that we observe the maximum possible synchrotron
power (at $1/\Gamma$).

Assuming steady state, we also have that the observed luminosity 
$L_{r,obs}$ (i.e. the sum of the synchrotron and the Compton luminosities)
integrated over the solid angle, cannot be greater than $L_{jet}$, i.e.
$\Gamma^2 L_r^\prime =L_{r,obs}/\Gamma^2 \le L_{jet}$.
Since $L_r^\prime$ depends on $<\gamma^2>$ which in turn is a function
of $\gamma_{max}$, we can limit $\gamma_{max}$ for a given  magnetic 
field and particle density (i.e. for a given $L_{jet}$).

\begin{figure}
\vskip -1 true cm
\psfig{file=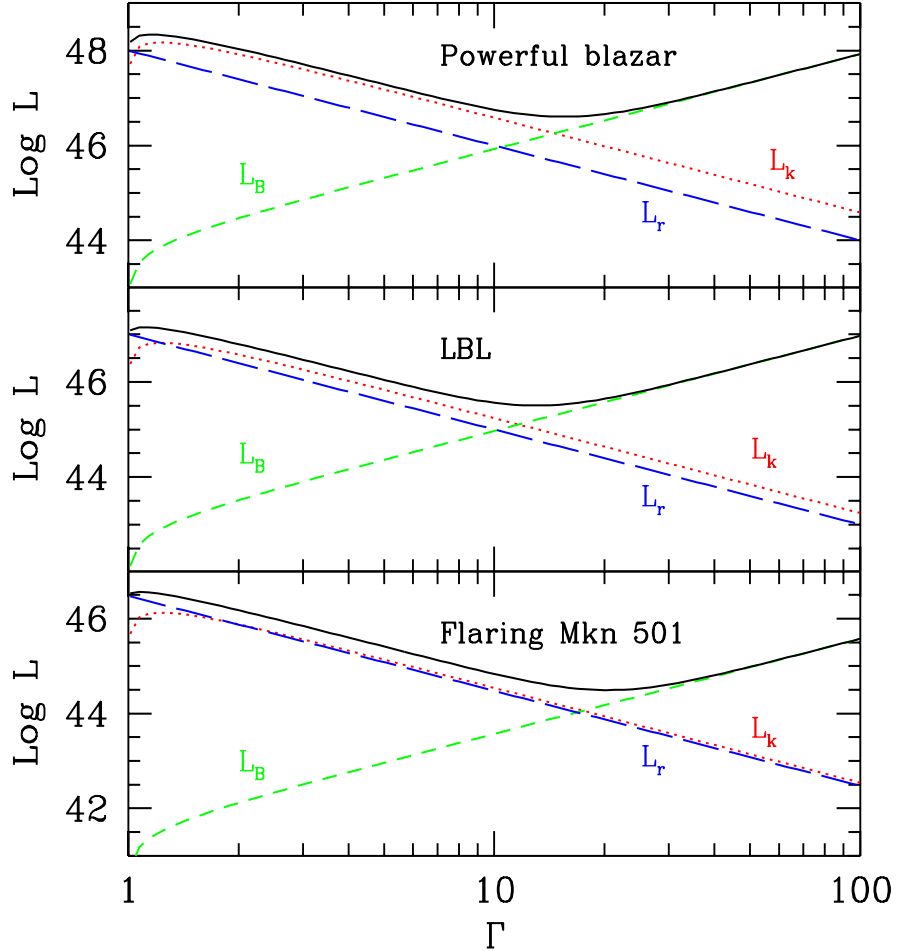,width=14.5truecm,height=15.5truecm}
\vskip -1 true cm
\caption[h]{The power of jets, in the form of radiation, bulk motion and
Poynting flux, as a function of bulk Lorentz factor.
Upper panel: $R=5\times 10^{16}$ cm, 
$B=3$ Gauss, $L_{s,obs}=10^{47}$ erg s$^{-1}$, 
$L_{r,obs}=10^{48}$ erg s$^{-1}$, $(<\gamma>+m_p/m_e)/<\gamma^2>=1$.
Mid panel: $R=5\times 10^{16}$ cm, 
$B=1$ Gauss, $L_{s,obs}=5\times 10^{46}$ erg s$^{-1}$, 
$L_{r,obs}=10^{47}$ erg s$^{-1}$, $(<\gamma>+m_p/m_e)/<\gamma^2>=10^{-2}$.
Bottom panel: $R=10^{16}$ cm,
$B=1$ Gauss, $L_{s,obs}=2\times 10^{46}$ erg s$^{-1}$, 
$L_{r,obs}=3\times 10^{46}$ erg s$^{-1}$, 
$(<\gamma>+m_p/m_e)/<\gamma^2>=10^{-3}$.
These parameters have been used after the fits to all EGRET blazars,
presented in Ghisellini et al. (1998) (upper two panels) and the model
used in Pian et al. (1998) for the flaring state of Mkn 501 of April 1997.
Note that for these parameters the bulk kinetic power and the Poynting flux
power are approximately equal for $\Gamma\sim$10--15, which is the value 
used for the fits. This strongly suggests that blazars jets are in
equipartition. From Ghisellini \& Celotti (1998).
}
\end{figure}

It is intriguing to interpret the Mkn 501 flare of April 1997, 
when BeppoSAX observed the peak of the synchrotron spectrum at 100 keV, 
by assuming that the available jet power was the same as in the quiescent 
state, while the efficiency of the radiative power to tap the available
energy increased because $\gamma_{max}\propto <\gamma^2>$ increased.
From Fig. 3 we then have that Mkn 501 was using, during the flare,
the entire jet power.
A further increase in $\gamma_{max}$ was therefore not possible,
since it would have resulted in $L_r^\prime\Gamma^2  > L_{jet}$.

Should other sources exist with even higher $\gamma_{max}$ than
Mkn 501 during flares?
The answer is yes, provided that the power balance is not violated:

\begin{equation}
\Gamma^2L_s^\prime \, <\, \Gamma^2 L_{r}^\prime\, <\,  L_{jet} \, \to \, 
{<\gamma^2> \over <\gamma>+m_p/m_e}  \, < \, 
{9 \pi \Gamma^2\over 128} \, {R m_ec^3 \over \sigma_T L_B}
\end{equation}
We see that larger values of $\gamma_{max}\propto <\gamma^2>$
are possible for smaller $L_B\sim L_{jet} \propto L_{s,obs}$.
This can be understood in this way:
the cooling rate for electrons of the highest energies is proportional
to $\gamma_{max}^2U_B$. 
At the limit this quantity is fixed by the total amount of
power that can be emitted.
Therefore $\gamma_{max}\propto 1/U_B \propto 1/L_B \propto 1/L_{s,obs}$.
It is then no coincidence that the 3 already known TeV sources
are three nearby, low luminosity BL Lacs.
More extreme sources (with larger $\gamma_{max}$) should be even
less powerful, and $weak$ TeV emitters, since the Klein Nishina
decline of the cross section becomes increasingly effective as
$\gamma_{max}$ increases.

\section{Matter content of AGN inner jets }

To summarize the findings about the power in bulk motion of AGN jets:
\begin{itemize}
\item There are different ways to measure the jet power.
One is to divide the energy content of the lobes by their life-time.
Another is to estimate the size, the particle density and the bulk
Lorentz factor at different locations in the jet.
The main uncertainty is to correctly calculate
the particle density, which depends on the low energy cut-off
of their energy distribution, not directly observable.
However, by assuming $\gamma_{min}\sim$30 and a proton--electron jet, 
the bulk motion jet power both at the VLBI scale and at the 
$\gamma$--ray emitting region scale are of the same order of the 
power required by the lobes to exist.

\item The jet power is also of the same order than the power in the 
radiation emitted by the accretion disk, illuminating the line emitting 
regions.
This may indicate the key role played by the magnetic field both
in extracting the gravitational energy of the accretion disk and
the rotational energy of the black hole.

\item Blazars work nearly at equipartition of $powers$.
This means that that they are very efficient synchrotron radiators.
The method of minimizing the sum of the bulk motion power in particles,
in Poynting flux and emitted radiation gives a particular value
of $\Gamma B$.
More importantly, we can set a lower limit to the jet power of AGNs.
\end{itemize}

With these informations we can try to work out some constraints
on the matter content of the inner part of jets, and the main
carriers of the jet power.
Previous reviews on this subject are by Celotti (1997, 1998),
and by Celotti \& Ghisellini (1998, in preparation).

Let assume that the central power house has to produce
a jet power $L_{jet}=10^{46}L_{jet,46}$ erg s$^{-1}$, and let us consider
different possibilities of the composition of jets at a distance
$z=10^{15} z_{15}$ cm from the black hole, and assume there
a cross section $R=10^{14} R_{14}$ cm of the jet.
The bulk  Lorentz factor is $\Gamma=10\Gamma_1$.
\vskip 0.5 true cm
\noindent
{\it  $\bullet$ Cold electron--positron pairs}
\vskip 0.2 true cm
\noindent
We can estimate how many pairs we need to carry $L_{jet}$, and
their corresponding scattering optical depth $\tau_\pm\equiv \sigma_TnR$,
where $n^\prime=n/\Gamma$:

\begin{equation}
\tau_{\pm} \, \sim\, 86 \, {L_{jet, 46} \over R_{14} \Gamma_1}
\end{equation}
With such a large optical depth, the pairs annihilate in a timescale
$\sim R/(c\tau_{\pm}) \ll R/c$, and the entire $L_{jet}$ is
transformed in a beamed annihilation line.
While annihilating, pairs will also scatter ambient photons, mainly
the UV produced in the inner accretion disk, producing, by bulk
Compton, a feature at $\sim \Gamma_1^2$ keV, of huge luminosity,
never observed.
We conclude that cold pairs cannot be dynamically important.

\vskip 0.3 true cm
\noindent
{\it $\bullet$ Hot electron--positron pairs}
\vskip 0.2 true cm

\noindent
If pairs are hot, with an average energy $<\gamma> m_ec^2$, 
their annihilation cross section is smaller, 
and fewer of them are required to carry $L_{jet}$.
But in this case a severe limit comes from the Compton emission
they would produce. Their Comptonization parameter $y$ is
(assuming $\tau_\pm<1$)

\begin{equation}
y_{\pm} \, \equiv\, \tau_{\pm} \Gamma^2 <\gamma^2> \,
\sim\, 8.6\times 10^3 \, {\Gamma_1 <\gamma^2> L_{jet, 46} 
\over R_{14} <\gamma>}
\end{equation}
which is huge. 
The cooling lifetime of these pairs is very short, resulting again
in a very rapid transformation of the entire $L_{jet}$ into 
Compton radiation.
Therefore even hot pairs cannot be dynamically important.

\vskip 0.3 true cm
\noindent
{\it $\bullet$ Cold Protons}
\vskip 0.2 true cm

\noindent
The scattering optical depth $\tau_p$ of the associated electrons
is a factor $m_p/m_e$ smaller, and the corresponding $y$ parameter,
assuming that these electrons are cold, is

\begin{equation}
y \, \sim \, 5 \, {\Gamma_1  L_{jet, 46} 
\over R_{14} }
\end{equation}
Bulk Compton radiation (see Sikora et al. 1997) 
should therefore be important, but the details of this signature
have still to be worked out, especially in the case of accelerating
plasma.
A detection of a feature resembling a `line' at 
$\nu\sim \Gamma^2 \nu_{UV} \sim \Gamma_1^2$ keV 
would be extremely important, revealing
both the value of the terminal Lorentz factor, the value of $L_{jet}$
and the fact that already in the inner regions the jet power is
transported in the form of cold protons.
It may be possible that the recently detected $\sim$1 kev feature
in PKS 0637--752 (Yakoob et al. 1998) is just this `line'.

\vskip 0.3 true cm
\noindent
{\it $\bullet$ Hot Protons}
\vskip 0.2 true cm

\noindent
If the protons have a mean Lorentz factor $<\gamma_p>$, we need
a factor $1/<\gamma_p>$ less protons to carry $L_{jet}$, thus
lowering the $y$ parameter by the same amount.
On the other hand, the value of the magnetic field necessary to confine 
them is large, comparable to the one necessary to transport $L_{jet}$
in the form of a Poynting flux, see below.

\vskip 0.3 true cm
\noindent
{\it $\bullet$ Poynting flux}
\vskip 0.2 true cm

\noindent
The value of the magnetic field corresponding to $L_{jet}\sim L_B$ is

\begin{equation}
B \, \sim \,  2\times 10^3 \, {L_{jet, 46}^{1/2} \over \Gamma_1 R_{14} }
\end{equation}
which is approximately the same value of the magnetic field
derived by $U_B\sim U_{r, disk}$ assuming $L_{disk}\sim L_{jet}$.

From the above simple estimates we conclude that, initially, $L_{jet}$
can be in the form of a Poynting flux, gradually accelerating matter 
until equipartition is reached, as pointed out in the 
studies of Li, Chiueh \& Begelman (1992) and Begelman \& Li (1994).

\section{ Galactic superluminals}
GRS 1915+105 and J1655--40 showed superluminal motion of both
approaching and receding blobs in radio interferometry observations
(Mirabel \& Rodriguez, 1994; Tingay et al. 1995; Hjellming \& Rupen 1995).
This allowed the determination of both the value of the true velocity
($\beta=0.92$ for both, corresponding to $\Gamma=2.55$) and the viewing
angle ($\theta\sim 70^\circ$ for GRS 1915+105 and slightly more
for J1655--40).
Their vicinity make it possible to detect their proper motion
with VLA, at sub-arcsecond resolution.
In Fig. 4 we present the SED of the two sources, from a collection
of data (of different epochs) taken from the literature.

The extremely complex variability of the X--ray emission is
currently monitored by RXTE, and is the subject of intense studies.
As Belloni et al. (1997) suggested, the X--ray data may indicate 
phases of accretion disk disruption and subsequent `re--filling',
also suggested by combined multiband optical and X--ray observations
in J1655-40 (Orosz et al. 1997).

\begin{figure}
\vskip -1.5 true cm
\psfig{file=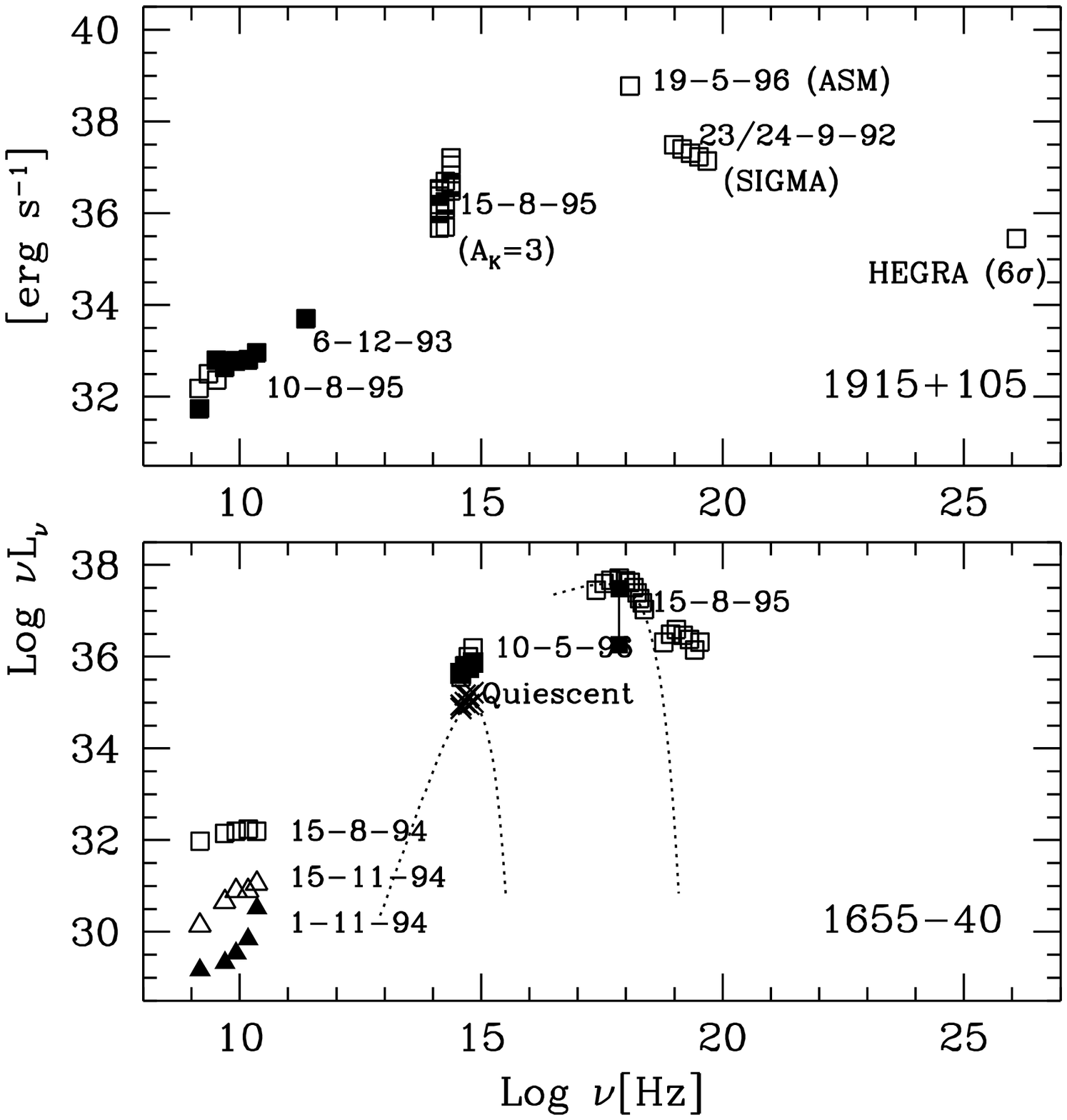,width=13 truecm,height=11truecm}
\vskip -.4 true cm
\caption[h]{The SED of GRS 1915+105 and J1655--40 constructed using
non--simultaneous data, taken from the literature.
Note, for 1915+105, the $6\sigma$ TeV value, obtained by HEGRA, and
reported in Aharonian \& Heinzelmann (1997). Dashed lines corresponds to 
the probable
spectrum of the companion star of 1655--40 and to a multi--color blackbody
spectrum, to represent its accretion disk emission.}
%
\psfig{file=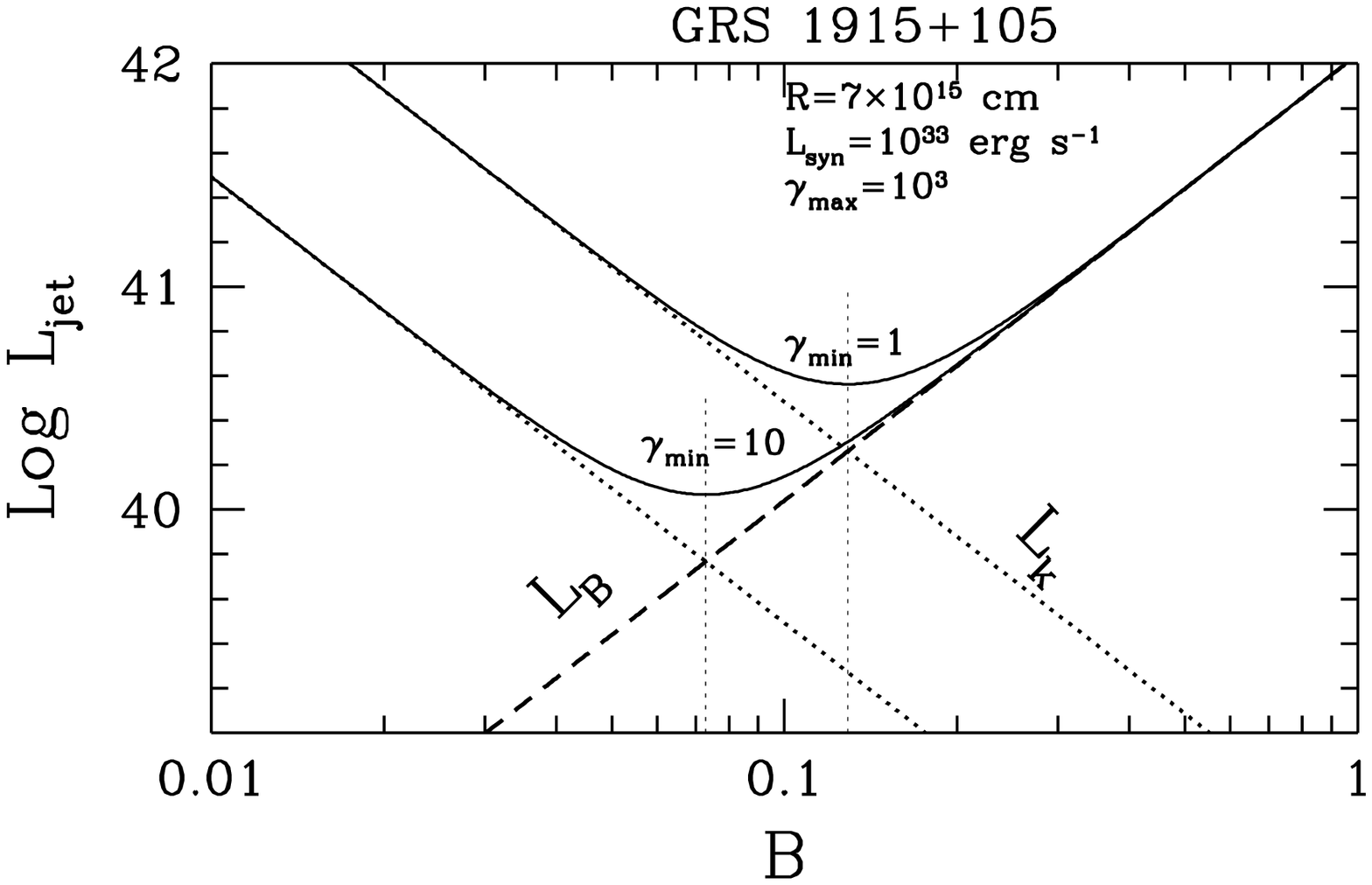,width=13 truecm,height=10 truecm}
\vskip -4 true cm
\caption[h]{The bulk motion power, the power in Poynting
flux and their sum as a function of the magnetic field
for GRS 1915+105. 
The adopted observed synchrotron luminosity corresponds to the radio
power of the April 1994 flare event.
It is assumed that the electron distribution is a power law,
$\propto \gamma^{-2}$ extending from $\gamma_{min}$ to $\gamma_{max}=10^3$.}
\end{figure}

We can repeat for GRS 1915+105 the same arguments we used for AGN,
trying to constrain the kinetic bulk power of its jets.
Assuming a size of $R=7\times 10^{15}$ cm emitting a synchrotron
observed luminosity of $L_{syn, obs}=10^{33}$ erg s$^{-1}$,
Gliozzi et al. (1998) derived, by the same method outlined 
in \S 6, a minimum jet power of
$L_{jet, min}\sim 10^{40}$ erg s$^{-1}$ for a low energy cut-off
in the electron distribution of $\gamma_{min}=10$, and assuming
an electron--proton jet
(these authors discuss and reject the possibility of a jet made
by electron--positron pairs).
Since in this case we know the bulk Lorentz factor and the viewing angle,
we also know the beaming factor, and the minimization of the total
jet power is for a particular value of the magnetic field $B$, as
illustrated in Fig. 5.

\section{Matter content of galactic superluminal jets}
The very large bulk kinetic power derived for GRS 1915+105
exceeds the power observed in X--rays, presumably
produced in the accretion disk, by a factor of a few.
We then derive

\begin{equation}
\dot M_{out} \, =\, 0.5  \, \left({\eta \over 0.1}\right)\, 
\left({2.55 \over \Gamma}\right)\, 
\left({ L_k \over 10^{40} {\rm erg~s^{-1} }}\right) 
\left({10^{39}{ \rm erg~s^{-1} } \over L_{disk}}\right) \dot M_{in} 
\end{equation}
This suggests that, if the jet material comes from the accretion disk,
it is a significant fraction of mass inflow rate, possibly corresponding
to `traumatic' disk changes.
On the other hand, it may well be that the above equations flags the
need for lower efficiencies $\eta$, to reduce the ratio
$\dot M_{out}/\dot M_{in}$ to the 1 per cent level, as in AGN.

We can here repeat the arguments discussed in \S 7 for the matter content
in the jets of GRS 1915+105, keeping in mind that in this case
we know $\Gamma$ and the viewing angle, corresponding
to a beaming factor $\delta\sim 0.5$.
A more detailed discussion can be found in Gliozzi et al. (1998).

\vskip 0.3 true cm

{\it $e^\pm$ pairs ---} If the jet is made by pairs at its start,
their density correspond to a very large scattering optical depth
[$\tau_\pm \sim 10^3 L_{k,40}/(\Gamma R_7)$], and
they cannot survive annihilation.
If the pairs are hot, with an average energy $<\gamma>m_ec^2$ at the start
of the jet, where they are embedded in a dense radiation field,
they cool very rapidly, converting a large fraction of $L_k$ into
radiation.
Then a pure pair plasma, neither cold or hot, can carry the kinetic power.

\vskip 0.3 true cm
{\it Electron -- proton plasma ---} If the jet is composed, initially,
by cold protons and hot electrons, with mean energy $<\gamma> m_e > m_p$,
we have the same case as above: the electrons rapidly cool.

There are no severe limits, instead, if both electrons and protons
are cold. 
The initial scattering optical depth is a few, and the bulk Compton
emission in our direction is at frequencies where the flux is dominated
by the radiation produced by the accretion disk.

If the protons are hot, the magnetic field needed to confine them
has a value of the same order than the magnetic field required to transport
$L_{jet}$ in the form of a Poynting vector.

\vskip 0.3 true cm
{\it Poynting vector ---} The magnetic field needed to carry $L_{jet}$
has a value
\begin{equation}
B\, \sim\, 3\times 10^8 \, {L_{jet,40}^{1/2} \over R_7}\,\,\, {\rm Gauss}
\end{equation}
This is also approximately the value needed to the Blandford--Znajek mechanism
to extract $10^{40}$ erg s$^{-1}$ from a maximally spinning black hole
of 10 solar masses.

\section{Conclusions}
We are starting to tackle the very important problem of how jets
are formed and accelerated by deriving their
energetics and matter content in different places along the jet 
and on the extended, isotropic lobes.

This was possible on one hand by having a sizeable sample of sources
observed with VLBI, and for which we know their apparent speeds,
and on the other hand by the discovery of $\gamma$--ray emission
of blazars, fixing the energetics of the observed radiation.
An important improvement in our knowledge would come from
observing features in the spectra of jet sources, originating
in the moving plasma, and flagging its bulk velocity.
One of them could be the bulk Compton `line' at $\sim$1keV
discussed by Sikora et al. (1997).
Other feature could be present, if the relativistic plasma
coexists with small blobs of cold materials, as suggested
by Celotti et al. (1998), and Celotti (1998, these proceedings).
Large area X--ray detectors, such as XMM and Constellation--X,
are probably the best instruments to find these important clues.

On a parallel path, the discovery of galactic superluminal sources
allows us to study nearby (bright) objects whose timescales,
scaling with the black hole mass, are a factor $\sim 10^8$ smaller
than what we observe in AGNs: a day in the lifetime of GRS 1915+105 
corresponds to $\sim 3\times 10^5$ years of a powerful blazar.
These `lab' systems will hopefully help us to understand
the jet phenomenon, and the strong gravity physics behind it.

\acknowledgments
It is a pleasure to thank Annalisa Celotti for years of discussions
and fruitful collaboration, and in particular for this review, largely
based on the very intense brainstorm we had in Graftavallen.


\begin{references}
\reference Aharonian F. \& Heinzelmann G., 1997, astro-ph/9702059
\reference Begelman M.C. \& Li Z.-Y., 1994, ApJ, 426, 269 
\reference Begelman M.C., Blandford R.D. \& Rees M.J., 1984,
      Rev. Mod. Phys., 56, 255 
\reference Belloni T., Mendez M., King A.R., van der Klis M. 
      \& van Paradijs J., 1997, ApJ 479, 145
\reference Blandford R.D. \& Znajek R.L., 1977, MNRAS, 176, 465
\reference Bloom S.D. et al. 1997, ApJ, 490, L145
\reference Buckley J.H. et al., 1996, ApJ 472, L9
\reference Celotti A., 1997, in Relativistic jets in AGNs,
      eds. M. Ostrowski, M. Sikora, G. Madejski \& M. Begelman,
      p. 270
\reference Celotti A., 1998, in Astrophysical jets: open problems,
      (Gordon \& Breach Science publ.), eds. S. Massaglia \& G. Bodo
      (Amsterdam), p. 79
\reference Celotti A. \& Fabian A.C. 1993, MNRAS, 264, 228
\reference Celotti A., Kuncic Z., Rees M.J. \& Wardle J.F.C. 1998, MNRAS, 
      293, 288
\reference Celotti A., Padovani P. \& Ghisellini G., 1997, MNRAS, 286, 415
\reference Dondi L. \& Ghisellini G., 1995, MNRAS, 273, 583
\reference Gaidos J.A. et al., 1996, Nature, 383, 319
\reference Ghisellini G., Celotti A., Fossati G., Maraschi L. \& Maraschi L.,
      1998, MNRAS, in press, astro-ph/9807317 
\reference Ghisellini G. \& Madau P., 1996, MNRAS, 280, 67
\reference Ghisellini G. \& Celotti A., 1998, submitted to MNRAS
\reference Ghisellini G., Padovani P., Celotti A. \& Maraschi L., 
      1993, ApJ, 407, 65
\reference Ghisellini G., Celotti A., George I.M. \& Fabian A.C., 1992,
      MNRAS, 258, 776 
\reference Gliozzi M., Bodo G. \& Ghisellini G., 1998, MNRAS, in press
\reference Hartman, R.C. et al., 1996, ApJ, 461, 698
\reference Hjellming R.M. \& Rupen M.P., 1995, Nature, 365, 464
\reference Li Z.-Y., Chiueh T. \& Begelman M.C., 1992, ApJ, 394, 459
\reference Macomb D.J. et al. 1995, ApJ, 449, L99
\reference Maraschi L., Ghisellini G., Tanzi E.G. \& Treves A., 1986, 
      ApJ, 310, 325
\reference Maraschi L., et al., 1994, ApJ, 435, L91
\reference Mirabel I.F. \& Rodriguez L.F., 1994, Nature, 371, 46
\reference Orosz J.A., Remillard R.A., Bailyn C.D. \& McClintock J.E., 
     1997, ApJ, 478, L83
\reference Pian E. et al., 1998, ApJ, 491, L17
\reference Rawlings S.G. \& Saunders R.D.E., 1991, Nature, 349, 138
\reference Rees M.J., 1996, Nature, 211, 468
\reference Sambruna R.M. et al., 1998, submitted to ApJ
\reference Sikora M., Madejski G., Mederski R. \&
      Poutanen J., 1997, ApJ, 484, 108
\reference Tingay S.J. et al., 1995 Nature, 374, 141
\reference Yakoob T., George I.M., Turner T.J., Nandra K., Ptak A.
      \& Serlemitsos P.J., 1998, ApJ in press (astro-ph/9807349)
\reference Wardle, J.F.C., 1977, Nature, 269, 563
\reference Wehrle A.E. et al., 1998, ApJ, 497, 178
\end{references}
\end{document}